\documentclass[journal]{IEEEtran}
\usepackage{ifpdf}

\usepackage{cite}

\ifCLASSINFOpdf
\else
\fi
\usepackage{graphicx,amsmath}
\usepackage{amsfonts}
\usepackage{algpseudocode}

\usepackage{booktabs}
\usepackage{multirow}
\usepackage[table]{xcolor}

\usepackage{flushend}

\usepackage[affil-it]{authblk}
\usepackage{hyperref}
\usepackage{url}

\newcommand{\secref}[1]{Section~\ref{#1}}
\newcommand{\figref}[1]{Fig.~\ref{#1}}
\newcommand{\tableref}[1]{Table~\ref{#1}}
\let\originaleqref=\eqref
\renewcommand{\eqref}{Eq.~\originaleqref}

\newcommand\correspondingauthor{\thanks{$^*$Corresponding author.}}

\begin{document}
%
\title{Polyphonic Sound Event Detection \\ by using Capsule Neural Networks}
%
%
%

\author{Fabio~Vesperini$^*$\correspondingauthor,~\IEEEmembership{}Leonardo~Gabrielli,~\IEEEmembership{}Emanuele~Principi,~\IEEEmembership{}and~Stefano~Squartini,~\IEEEmembership{Senior Member,~IEEE}
\thanks{The authors are with the A3lab, Department of Information Engineering,

	Universit\`a Politecnica delle Marche, Ancona (Italy), E-mail: 
	
	f.vesperini@pm.univpm.it, \{l.gabrielli,e.principi,s.squartini\}@univpm.it}}

%
%

\markboth{Journal of Selected Topics in Signal Processing,~Vol.~X, No.~X, OCTOBER~2018}%
{Vesperini {\textit{et al.}}: Polyphonic SED using CapsNet}
%




\maketitle

\begin{abstract}

Artificial sound event detection (SED) has the aim to mimic the human ability to perceive and understand what is happening in the surroundings. 
\textcolor{black}{Nowadays, learning offers valuable techniques for this goal such as convolutional neural networks (CNNs). 
The capsule neural network (CapsNet)} architecture has been recently introduced in the image processing field with the intent to overcome some of the known limitations of CNNs, specifically regarding the scarce robustness to affine transformations (i.e., perspective, size, orientation) and the detection of overlapped images.
This motivated the authors to employ CapsNets to deal with the polyphonic SED task, in which multiple sound events occur simultaneously.
Specifically, we propose to exploit the capsule units to represent a set of distinctive properties for each individual sound
event. Capsule units are connected through a so-called \textit{dynamic routing} that encourages learning part-whole relationships and improves the detection performance in a polyphonic context.
This paper reports extensive evaluations carried out on three publicly available datasets, 
showing how the CapsNet-based algorithm not only outperforms standard CNNs but also allows to achieve the best results with respect to the state-of-the-art algorithms.
\end{abstract}


%
\IEEEpeerreviewmaketitle

\section{Introduction}
%
%
%
%
\IEEEPARstart{H}{uman} cognition relies on the ability to sense, process, and understand the surrounding environment and its sounds.
Although the skill of listening and understanding their origin is so natural for living beings, it still results in a very challenging task for computers. 

Sound event detection (SED), or acoustic event detection, has the aim to mimic this cognitive feature by means of artificial systems. Basically, a SED algorithm is designed to detect the onset and offset times for a variety of sound events captured in an audio recording and associate a textual descriptor, i.e., a label for each of these events.

In recent years, SED has received \textcolor{black}{significant interest} from the computational auditory scene analysis community \cite{virtanen2018computational}, due to its potential in several engineering applications. 
Indeed, the automatic recognition of sound events and scenes can have a considerable impact in a wide range of applications where sound or sound sensing is advantageous with respect to other modalities. 
This is the case of acoustic surveillance \cite{crocco2016audio}, healthcare monitoring \cite{peng2009healthcare, foggia2015reliable} or urban sound analysis \cite{salamon2017deep}, where the short duration of certain events (i.e., a human fall, a gunshot or a glass breaking) or the personal privacy motivate the exploitation of \textcolor{black}{audio information} rather than, e.g., \textcolor{black}{image processing.}
In addition, audio processing is often less computationally demanding compared to other multimedia domains, 
thus embedded devices can be easily equipped with microphones and sufficient computational capacity to locally process the signal captured. 
These could be smart home devices for home automation purposes or sensors for wildlife and biodiversity monitoring (i.e., bird calls detection \cite{grill2017two}).

SED algorithms in a real-life scenario face many challenges. These include the presence of simultaneous events, environmental noise and events of the same class produced by different sources \cite{stowell2015acoustic}. Since multiple events are very likely to overlap, a \textit{polyphonic} SED algorithm, i.e., an algorithm able to detect multiple simultaneous events, needs to be designed.
Finally, the effects of noise and intra-class variability represent further challenges for SED in real-life situations. 

Traditionally, \textcolor{black}{polyphonic acoustic event analysis} has been approached with statistical modelling methods, including hidden Markov models (HMM) \cite{degara2011onset}, Gaussian mixture models (GMM) \cite{heittola2010audio}, non-negative matrix Factorization (NMF) \cite{carabias2011musical} and support vector machines (SVM) \cite{guo2003content}. In the recent era of \textcolor{black}{``deep learning''}, different neural network architectures have been successfully used for sound event detection and classification tasks, including feed-forward neural networks (FNN) \cite{mcloughlin2015robust}, deep belief networks \cite{mohamed2012acoustic}, convolutional neural networks (CNNs) \cite{piczak2015environmental} and recurrent neural networks (RNNs) \cite{graves2013speech}. In addition, these architectures laid the foundation for end-to-end systems \cite{trigeorgis2016adieu, wu2017end}, in which the feature representation of the audio input is automatically learnt from the raw audio signal waveforms. 

\subsection{Related Works}

The use of deep learning models has been motivated by the increased availability of datasets and computational resources and resulted in significant performance improvements. 
The methods based on CNNs and RNNs have established the new state-of-the-art performance on the SED task, thanks to the capabilities to learn the non-linear relationship between time-frequency features of the audio signal and a target vector representing sound events. In \cite{espi2015}, the authors show how ``local'' patterns can be learned by a CNN and can be exploited to improve the performance of detection and classification of non-speech acoustic events occurring in conversation scenes, in particular compared to a FNN-based system which processes multiple resolution spectrograms in parallel. 

The combination of the CNN structure with recurrent units has increased the detection performance by taking advantage of the characteristics of each architecture. This is the case of convolutional recurrent neural networks (CRNNs) \cite{cakir2017convolutional}, which provided state-of-the-art performance especially in the case of polyphonic SED. CRNNs consolidate the CNN property of local shift invariance with the capability to model short- and long-term temporal dependencies provided by the RNN layers. This architecture has been also employed in almost all of the most performing algorithms proposed in the recent editions of research challenges such as the IEEE Audio and Acoustic Signal Processing (AASP) Challenge on  Detection and Classification of Acoustic Scenes and Events (DCASE) \cite{DCASE2017Workshop}. On the other hand, if the datasets are not sufficiently large, problems such as overfitting can be encountered with these models, which typically are composed of a considerable number of free-parameters (i.e., more than 1M). 

Encouraging polyphonic SED performance has been obtained using CapsNets in preliminary experiments conducted on the Bird Audio Detection task in occasion of the DCASE 2018 challenge \cite{vesperini2018capsule}, confirmed by the results reported in \cite{iqbal2018capsule}.
The CapsNet \cite{sabour2017dynamic} is a recently proposed architecture for image classification and it is based on the grouping of activation units into novel structures introduced in \cite{hinton2011transforming}, named \textit{capsules}, along with a procedure called dynamic routing. The capsule has been designed to represent a set of properties for an entity of interest, while dynamic routing is included to allow the network to implicitly learn global coherence and to identify part-whole relationships between capsules.

The authors of \cite{sabour2017dynamic} show that CapsNets outperform state-of-the-art approaches based on CNNs for digit recognition in the MNIST dataset case study.
They designed the CapsNet to learn how to assign the suited partial information to the entities that the neural network has to predict in the final classification. This property should overcome the limitations of solutions such as max-pooling, currently employed in CNNs to provide local translation invariance, but often reported to cause an excessive information loss. Theoretically, the introduction of the dynamic routing can supply invariances for any property captured by a capsule, allowing also to adequately train the model without requiring extensive data augmentation or dedicated domain adaptation procedures.

\subsection{Contribution}

The proposed system is a fully data-driven approach based on the CapsNet deep neural architecture presented by Sabour et al.\ \cite{sabour2017dynamic}. This architecture has shown promising results on the classification of highly overlapped digit images. In the audio field, a similar condition can be found in the detection of multiple concomitant sound events from acoustic spectral representations, thereby we propose to employ the CapsNet for \textcolor{black}{polyphonic SED} in real-life recordings.
The novel computational structure based on capsules, combined with the routing mechanism, allows to be invariant to intra-class affine transformations and to identify part-whole relationships between data features. In the SED case study, it is hypothesized that this characteristic confers to CapsNet the ability to effectively select most representative spectral features of each individual sound event and separate them from overlapped descriptions of the other sounds in the mixture. 

This hypothesis is supported by previously mentioned related works. Specifically, in \cite{vesperini2018capsule}, the CapsNet is exploited in order to obtain the prediction of the presence of heterogeneous polyphonic sounds (i.e., bird calls) on unseen audio files recorded in various conditions. \textcolor{black}{In \cite{iqbal2018capsule}, the authors proposed a CapsNet for sound event detection that uses gated convolutions in the initial layers of the network, and an attention layer that operates in parallel with the final capsule layer. The outputs of the two layers are merged and used to obtain the final prediction. The algorithm is evaluated on the weakly-labeled dataset of the DCASE 2017 challenge \cite{DCASE2017challenge} with promising results. In \cite{bae2018}, capsule networks have been applied to a speech command recognition task, and the authors obtained a significant performance improvement with respect to CNNs.}


In this paper, we present an extensive analysis of SED conducted on real-life audio datasets and compare the results with state-of-the-art methods. 
In addition, we propose a variant of the dynamic routing procedure which takes into account the temporal dependence of adjacent frames. 
The proposed method outperforms previous SED approaches in terms of detection error rate in the case of polyphonic SED, while it has comparable performance with respect to CNNs in the case of monophonic SED. 

The whole system is composed of a feature extraction stage and a detection stage. The feature extraction stage transforms the audio signal into acoustic spectral features, while the second stage processes these features to detect the onset and offset times of specific sound events.
In this latter stage we include the capsule units. The network parameters are obtained by supervised learning using annotations of sound events activity as target vectors. We have evaluated the proposed method against three datasets of real-life recordings and we have compared its performance both with the results of experiments with a traditional CNN architecture, and with the performance of well-established algorithms which have been assessed on the same datasets.

The rest of the paper is organized as follows. In \secref{sec:proposed-meth} the task of polyphonic SED is formally described and the stages of the approach we propose are detailed, including a presentation of the CapsNet architecture characteristics. In \secref{sec:experiment}, we present the evaluation set-up used to accomplish the performance of the algorithm we propose and the comparative methods. In \secref{sec:results} the results of experiments are discussed and compared with baseline methods. \secref{sec:conclusions} finally presents our conclusions for this work.

\section{Proposed Method}
\label{sec:proposed-meth}

The aim of polyphonic SED is to find and classify the sound events present in an audio signal. The algorithm we propose is composed of two main stages: sound representation and polyphonic detection. In the sound representation stage, the audio signal is transformed in a two-dimensional time-frequency representation to obtain, for each frame $t$ of the audio signal, a feature vector $\mathbf{x}_t \in \mathbb{R}^F$, where $F$ represents the number of frequency bands. 

Sound events possess temporal characteristics that can be exploited for SED, thus certain events can be efficiently distinguished by their time evolution. Impulsive sounds are extremely compact in time (e.g., gunshot, object impact), while other sound events have indefinite length (i.e., wind blowing, people walking). Other events can be distinguished from their spectral evolution (e.g., bird singing, car passing by). Long-term time domain information is very beneficial for SED and motivates for the use of a temporal \textit{context} allowing the algorithm to extract information from a chronological sequence of input features. Consequently, these are presented as a context window matrix $\mathbf{X}_{t:t+T-1} \in \mathbb{R}^{T \times F \times C}$, where $T\in \mathbb{N}$ is the number of frames that defines the sequence length of the temporal context, $F\in \mathbb{N}$ is the number of frequency bands and $C$ is the number of audio channels. Differently, the target output matrix is defined as $\mathbf{Y}_{t:t+T-1} \in \mathbb{N}^{T \times K}$, where $K$ is the number of sound event classes.

In the SED stage, the task is to estimate the probabilities $p(\mathbf{Y}_{t:t+T-1}| \mathbf{X}_{t:t+T-1},  \boldsymbol{\theta}) \in \mathbb{R}^{T \times K}$, 
where $ \boldsymbol{\theta}$ denotes the parameters of the neural network.
The network outputs, i.e., the event activity probabilities, are then compared to a threshold in order to obtain event activity predictions $\mathbf{\hat{Y}}_{t:t+T-1}  \in \mathbb{N}^{T \times K}$.
The parameters $ \boldsymbol{\theta}$  are trained by supervised learning, using the frame-based annotation of the sound event class as target output, thus, if class $k$ is active during frame $t$, $Y(t,k)$ is equal to 1, and is set to 0 otherwise. The case of polyphonic SED implies that this target output matrix can have multiple non-zero elements $K$ in the same frame $t$, since several classes can be simultaneously present. 

Indeed, polyphonic SED can be formulated as a multi-label classification problem in which the sound event classes are detected by multi-label annotations over consecutive time frames. The onset and offset time for each sound event are obtained by combining the classification results over consequent time frames. The trained model will then be used to predict the activity of the sound event classes in an audio stream without any further post-processing operations and prior knowledge on the events locations.

\subsection{Feature Extraction}

For our purpose, we use two acoustic spectral representation, the magnitude of the Short Time Fourier Transform (STFT) and \textit{LogMel} coefficients, obtained from all the audio channels and extensively used for other SED algorithms. Except where differently stated, we study the performance of binaural audio features and compare it with those extracted from a single channel audio signal. In all cases, we operate with audio signals sampled at 16\,kHz and we calculate the STFT with a frame size equal to 40\,ms and a frame step equal to 20\,ms. Furthermore, the audio signals are normalized to the range $[-1, 1]$ in order to have the same dynamic range for all the recordings.

The STFT is computed on $1024$ points for each frame, while LogMel coefficients are obtained by filtering the STFT magnitude spectrum with a filter-bank composed of \textcolor{black}{40 triangular filters evenly spaced in the mel frequency scale \cite{slaney1998auditory}.}
In both cases, the logarithm of the energy of each frequency band is computed. 
The input matrix $\mathbf{X}_{t:t+T-1}$ concatenates $T=256$ consequent STFT or LogMel vectors for each channel $C=\{1,2\}$, thus the resulting feature tensor is $\mathbf{X}_{t:t+T-1} \in \mathbb{R}^{256\times F \times C}$, where $F$ is equal to 513 for the STFT and equal to 40 for the LogMels.
The range of feature values is then normalized according to the mean and the standard deviation computed on the training sets of the neural networks.

\subsection{\textcolor{black}{Background on capsule networks}}\label{ssec:CapsNet}
\textcolor{black}{
    Capsules have been introduced to overcome some limitations of CNNs, in particular the loss of information caused by the max-pooling operator used for obtaining translational invariance \cite{hinton2011transforming,sabour2017dynamic}. The main idea behind capsules is to replace conventional neurons with local units that produce a vector output (capsules) incorporating all the information detected in the input. Moreover, lower-level capsules are connected to higher-level ones with a set of weights determined during inference by using a dynamic routing mechanism. These two aspects represent the main differences from conventional neural networks, where neurons output a single scalar value, and connection weights are determined in the training phase by using back-propagation \cite{hinton2011transforming,sabour2017dynamic}.
}
{
Recalling the original formulation in \cite{hinton2011transforming,sabour2017dynamic}, a layer of a capsule network is divided in multiple computational units named capsules. Considering capsule $j$, its total input $\mathbf{s}_j$ is calculated as:
\begin{equation}\label{eq:single_caps}
\mathbf{s}_j = \sum_i \alpha_{ij} \mathbf{W}_{ij}\mathbf{u}_i = \sum_i \alpha_{ij} \hat{\mathbf{u}}_{j|i},
\end{equation}
where $\alpha_{ij}$ are coupling coefficients between capsule $i$ and capsule $j$ in the lower-level layer, $\mathbf{u}_i$ is the output of capsule $i$, $\mathbf{W}_{ij}$ are transformation matrices, and $\hat{\mathbf{u}}_{j|i}$ are prediction vectors. The vector output of  capsule $j$ is calculated by applying a non-linear squashing function that makes the length of short vectors close to zero and the length of long vectors close to 1:
\begin{equation}
\label{eq:squashing}
\mathbf{v}_j  = \frac{\| \mathbf{s}_j \|^2}{ 1 + \| \mathbf{s}_j \|^2} \frac{\mathbf{s}_j}{ \|\mathbf{s}_j \|}.
\end{equation}
Using the squashing function of \eqref{eq:squashing} allows to interpret the magnitude of the vector as a probability, in particular the probability that the entity represented by the capsule is present in the input \cite{sabour2017dynamic}.
}


\textcolor{black}{
Coefficients $\alpha_{ij}$ measure how likely capsule $i$ may activate capsule $j$. Thus, the value of $\alpha_{ij}$ should be relatively high if the properties of capsule $i$ coincide with the properties of capsule $j$ in the layer above. As shown in detail in the next section, this is obtained by using the notion of \textit{agreement} between capsules in two consecutive layers. The coupling coefficients are calculated by the iterative process of dynamic routing, and capsules in the higher layers should \textcolor{black}{include} capsules in the layer below in terms of the entity they identify. Dynamic routing iteratively attempts to find these associations and supports capsules to learn features that ensure these connections. The new ``routing-by-agreement'' algorithm introduced in \cite{sabour2017dynamic} represents an evolution of the simpler routing mechanism intrinsic in max-pooling and will be described in the next section.
}

\subsubsection{\textcolor{black}{Dynamic Routing}}
\label{ssec:routing}
\textcolor{black}{
After giving a qualitative description of the routing mechanism, we describe in detail the algorithm used in \cite{sabour2017dynamic} to compute the coupling coefficients.} 



\begin{figure}[t]
    \textcolor{black}{
\begin{algorithmic}[1]
    \Procedure{Routing}{$\hat{\mathbf{u}}_{ij}$, $r$, $l$}
    \State $\forall$ capsule $i$ in layer $l$ and capsule $j$ in layer $(l+1)$: $\beta_{ij}$ $\gets$ 0.
    \For {$r$ iterations}
        \State $\forall$ capsule $i$ in layer $l$: $\alpha_{ij} \gets$ $\frac{\exp(\beta_{ij})  }{\sum_k \exp(\beta_{ik})}$
        \State $\forall$ capsule $j$ in layer $(l+1)$: $\mathbf{s}_j\gets \sum_i\alpha_{ij}\hat{\mathbf{u}}_{j|i}$
        \State $\forall$ capsule $j$ in layer $(l+1)$: $\mathbf{v}_j \gets \frac{\| \mathbf{s}_j \|^2}{ 1 + \| \mathbf{s}_j \|^2} \frac{\mathbf{s}_j}{ \|\mathbf{s}_j \|}$
        \State $\forall$ capsule $i$ in layer $l$ and capsule $j$ in layer $(l+1)$: $\beta_{ij} \gets \beta_{ij}+\hat{\mathbf{u}}_{j|i}\cdot \mathbf{v}_j$
    \EndFor
    \State \textbf{return} $\mathbf{v}_j$
    \EndProcedure
\end{algorithmic}
\caption{The dynamic routing algorithm proposed in \cite{sabour2017dynamic}.}\label{fig:routing}
}
\end{figure}

\textcolor{black}{
The ``routing-by-agreement'' algorithm operates as shown in  \figref{fig:routing}. The algorithm is executed for each layer $l$ of the network and for $r$ iterations, and it outputs vectors $\mathbf{v}_j$ of layer $(l+1)$. In essence, the algorithm represents the forward pass of the network. As shown in line 4, coupling coefficients $\alpha_{ij}$ are determined by applying the softmax function to coefficients $\beta_{ij}$:
\begin{equation}
\alpha_{ij} = \frac{\exp(\beta_{ij})  }{\sum_k \exp(\beta_{ik})}.
\end{equation}
The softmax function ensures that $\alpha_{ij} \in (0,1)$, thus making $\alpha_{ij}$ the probability that capsule $i$ in the lower-level layer sends its output to capsule $j$ in the upper-level layer. The coefficients $\beta_{ij}$ are initialized to zero so that the coupling coefficients $\alpha_{ij}$ have all the same initial value. After this step, the $\beta_{ij}$ coefficients are updated by using an iterative algorithm which uses the \textit{agreement} between the output of capsule $j$, $\mathbf{v}_j$, and the prediction of capsule $i$, $\hat{\mathbf{u}}_{ij}$, in the layer below. The agreement is measured by the scalar product $\hat{\mathbf{u}}_{j|i} \cdot \mathbf{v}_j$, and it provides a measure of how similar the directions (i.e., the properties of the entity they represent) of capsules $i$ and $j$ are. 
}

\subsubsection{Margin loss function}
The length of the vector $\mathbf{v}_j$ is used to represent the probability that the entity represented by the capsule $j$ exists. The CapsNet have to be trained to produce long instantiation vector at the corresponding $k_{th}$ capsule if the event that it represents is present in the input audio sequence.
A separate margin loss is defined for each target class $k$ as:
\begin{equation}
\begin{split}
L_k = T_k \max(0, m^+ - & \left \|\mathbf{v}_j \right \|)^2 + \\
&\lambda(1 - T_k)\max(0, \left \|\mathbf{v}_j \right \| - m^-)^2,
\end{split}
\end{equation}
where $T_k = 1$ if an event of class $k$ is present, while $\lambda$ is a down-weighting factor of the loss for absent sound event classes. 
$m^+$, $ m^-$ and $\lambda$ are respectively set equal to 0.9, 0.1 and 0.5 as suggested in \cite{sabour2017dynamic}.
The total loss is simply the sum of the losses of all the output capsules.

\subsection{CapsNet for Polyphonic Sound Event Detection}
\begin{figure}[t]
    \centering
    \includegraphics[width=\columnwidth]{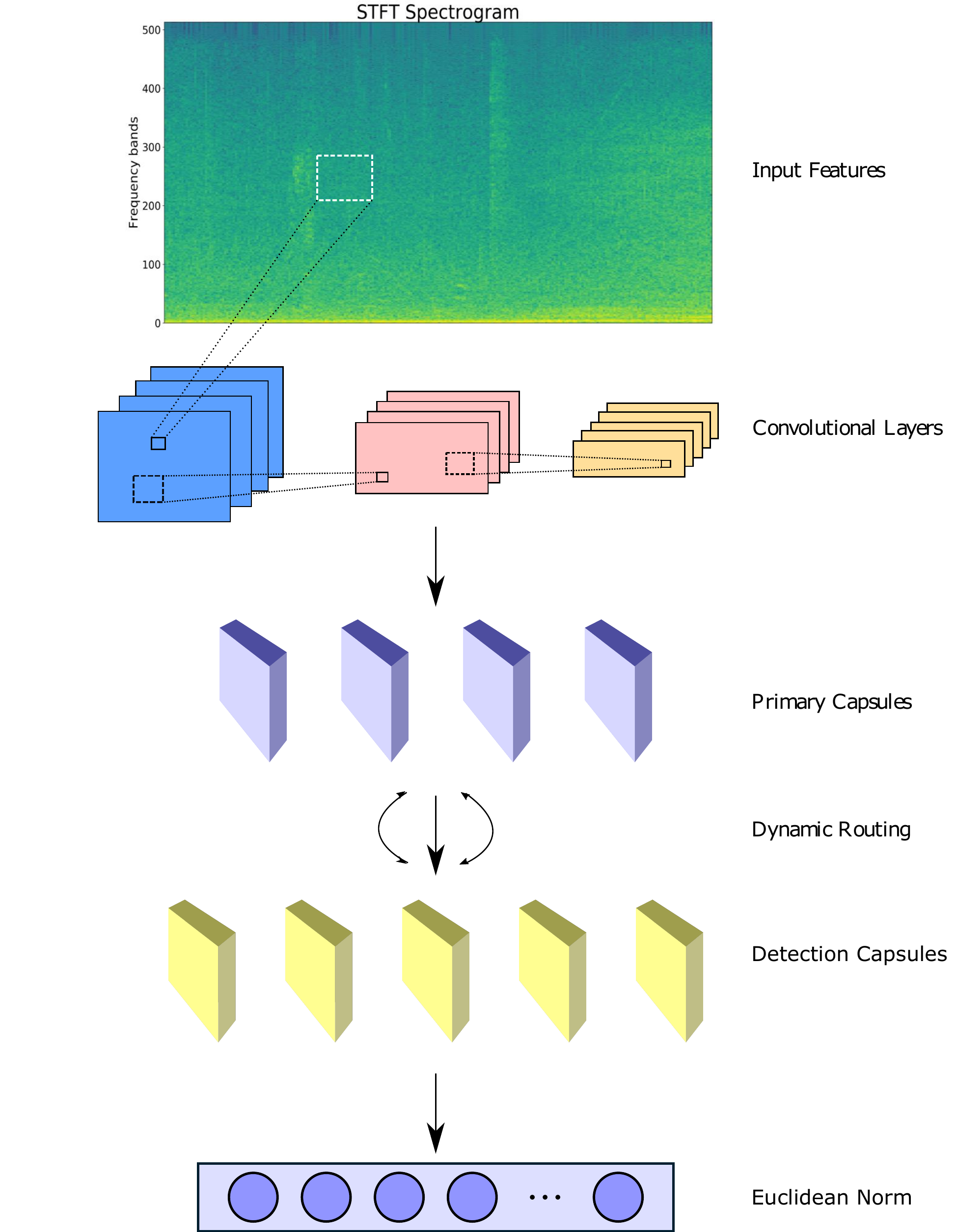}
    \caption{Flow chart of the capsule neural network architecture used for polyphonic sound event detection.}
    \label{fig:flowchart}
\end{figure}

\begin{figure*}[t]
    \centering
    \includegraphics[width=\textwidth]{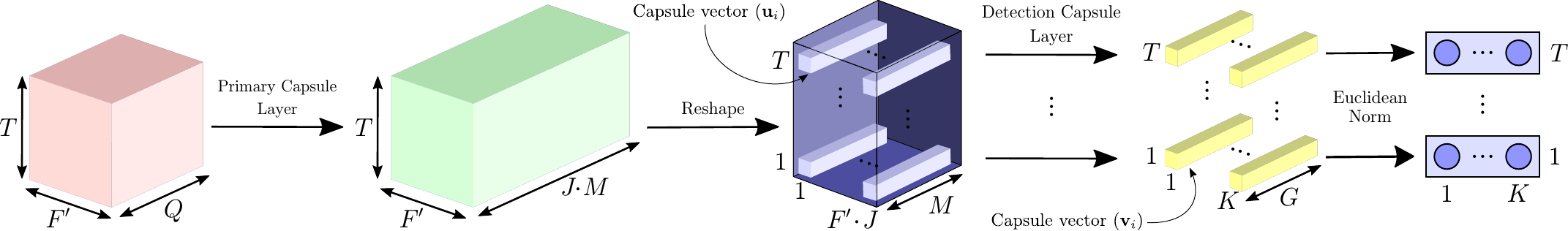}
    \caption{\textcolor{black}{Details of the processing stages that occur after the initial CNN layers. The dimension of vectors $\mathbf{u}_i$ is $1 \times 1 \times M$, the dimension of vectors $\mathbf{v}_j$ is $1 \times 1 \times G$. The decision stage after the Euclidean norm calculation is not shown for simplicity.}}
    \label{fig:dimensions}
\end{figure*}

The architecture of the neural network is shown in \figref{fig:flowchart}. The first stages of the model are traditional CNN blocks which act as feature extractors on the input $\mathbf{X}_{t:t+T-1}$. 
\textcolor{black}{The input of each CNN block is zero-padded in order to preserve its dimension, and, after each block, max-pooling \cite{scherer2010evaluation}} is used to halve the dimensions only on the frequency axis. \textcolor{black}{Thus, the output of the first CNN layers has dimension $T\times F' \times Q$, where $F'<F$ is the number of elements in the frequency axis after max-pooling, and $Q$ is the number of kernels in the final CNN block. This tensor is then used as input for the Primary Capsule Layer that represents the lowest level of multi-dimensional entities. The processing stages occurring after the CNN blocks are depicted in \figref{fig:dimensions}.} Basically, the Primary Capsule Layer is a convolutional layer with $J \cdot M$ filters, i.e., it contains $M$ convolutional capsules with $J$ kernels each. \textcolor{black}{The output tensor of this layer has dimension $T \times F' \times J\cdot M$, and it is then reshaped in order to obtain a $T \times F'\cdot J \times M$ tensor.  Capsule vectors $\mathbf{u}_i$ are represented by the $M$-dimensional $T \cdot F'\cdot J$ vectors of this tensor obtained after applying the squashing operation of \eqref{eq:squashing}.  The final layer, or Detection Capsule Layer, is a time-distributed layer composed of $K$ densely connected capsule units with $G$ elements. With ``time-distributed'', we mean that the same weights are applied for each time-index. For each $t$, thus, the Detection Capsule Layer outputs $K$ vectors $\mathbf{v}_i$ composed of $G$ elements. This differs from the architecture proposed in \cite{bae2018}, where all the capsule vectors from the Primary Capsule Layer are processed as a whole.} Since the previous layer is also a capsule layer, the dynamic routing algorithm is used to compute the output. The \textit{background} class was included in the set of $K$ target events, in order to represent its instance with a dedicated capsule unit and train the system to recognize the absence of events. In the evaluation, however, we consider only the outputs relative to the target sound events. The model predictions are obtained by computing the Euclidean norm of the output of each Detection Capsule. These values represent the probabilities that one of the target events is active in a frame $t$ of the input feature matrix $\mathbf{X}_{t:t+T-1}$, thus we consider them as the network output predictions.

In \cite{sabour2017dynamic}, the authors propose a series of densely connected neuron layers stacked at the bottom of the CapsNet, with the aim to regularize the weights training by reconstructing the input image. Here, this technique entails an excessive complexity of the model to train, due to the higher number of units needed to reconstruct $\mathbf{X}_{t:t+T-1} \in \mathbb{R}^{T \times F \times C}$, yielding poor performance in our preliminary experiments. We decided, thus, to use dropout \cite{srivastava2014dropout} and $L_2$ weight normalization \cite{hoerl1970ridge} as regularization techniques, as done in \cite{iqbal2018capsule}.

\section{Experimental Set-Up}
\label{sec:experiment}

In order to evaluate the performance of the proposed method, we performed a series of experiments on three datasets provided to the participants of different editions of the DCASE challenge \cite{DCASE2017challenge, mesaros2016tut}. We evaluated the results by comparing the system based on the Capsule architecture with the traditional CNN. The hyperparameters of each network have been optimized with a \textit{random search} strategy \cite{bergstra2012random}. Furthermore, we reported the baselines and the best state-of-the-art performance provided by the challenge organizers.

\subsection{Dataset}
\label{ssec:dataset}

We assessed the proposed method on three datasets, two containing stereo recordings from real-life environments and one artificially generated monophonic mixtures of isolated sound events and real background audio.

In order to evaluate the proposed method in polyphonic real-life conditions, we used the TUT Sound Events 2016 \& 2017 datasets, which were included in the corresponding editions of the DCASE Challenge. For the monophonic SED case study, we used the TUT Rare Sound Events 2017 which represents the task 2 of the DCASE 2017 Challenge.

\subsubsection{TUT Sound Events 2016}
The TUT Sound events 2016 (TUT-SED 2016)\footnote{\url{http://www.cs.tut.fi/sgn/arg/dcase2016/}} dataset consists of recordings from two acoustic scenes, respectively ``Home'' (indoor) and ``Residential area'' (outdoor) which we considered as two separate subsets. These acoustic scenes were selected from the challenge organizers to represent common environments of interest in applications for safety and surveillance (outside home) and human activity monitoring or home surveillance \cite{mesaros2016tut}.
\textcolor{black}{The dataset was collected in Finland by the Tampere University of Technology from different locations by means of a binaural recording system.} 
A total amount of around 54 and 59 minutes of audio are provided respectively for ``Home'' and ``Residential area'' scenarios.
Sound events present in each recording were manually annotated without any further cross-verification, due to the high level of subjectivity inherent to the problem. 
For the ``Home'' scenario a total of 11 classes were defined, 
while for the ``Residential Area'' scenario 7 classes were annotated. 

Each scenario of the TUT-SED 2016 has been divided into two subsets: Development dataset and Evaluation dataset. The split was done based on the number of examples available for each sound event class. In addition, for the Development dataset a cross-validation setup is provided in order to easily compare the results of different approaches on this dataset. The setup consists of 4 folds, so that each recording is used exactly once as test data. More in detail, the ``Residential area'' set consists of 5 recordings in the Evaluation set and 12 recordings in the Development set, while the ``Home'' set consists of 5 recordings in the Evaluation set and 10 recordings in turn divided into 4 folds as training and validation subsets.

\subsubsection{TUT Sound Events 2017}
The TUT Sound Events 2017 (TUT-SED 2017)\footnote{\label{note_dcase17}\url{http://www.cs.tut.fi/sgn/arg/dcase2017/}} dataset consists of recordings of street acoustic scenes with various levels of traffic and other activities, for a total of 121 minutes of audio. The scene was selected as representing an environment of interest for detection of sound events related to human activities and hazard situations. It is a subset of the TUT Acoustic scenes 2016 dataset \cite{mesaros2016tut}, from which also TUT-SED 2016 dataset was taken. Thus, the recording setup, the annotation procedure, the dataset splitting, and the cross-validation setup is the same described above. \textcolor{black}{They share also some audio contents, in particular the ``Residential area'' scenario.} The 6 target sound event classes were selected to represent common sounds related to human presence and traffic, and they include brakes squeaking, car, children, large vehicle, people speaking, people walking. The Evaluation set of the TUT-SED 2017 consists of 29 minutes of audio, whereas the Development set is composed of 92 minutes of audio which are employed in the cross-validation procedure.

\subsubsection{TUT Rare Sound Events 2017} 
The TUT Rare Sound Events 2017 (TUT-Rare 2017)\textsuperscript{\ref{note_dcase17}} \cite{DCASE2017challenge} consists of isolated sounds of three different target event classes (respectively, baby crying, glass breaking and gunshot) and 30-second long recordings of everyday acoustic scenes to serve as background, such as park, home, street, cafe, train, etc. \cite{mesaros2016tut}. In this case we consider a \textit{monophonic}-SED, since the sound events are artificially mixed with the background sequences without overlap. In addition, the event potentially present in each test file is known a-priori thus it is possible to train different models, each one specialized for a sound event. In the Development set, we used a number of sequences equal to 750, 750 and 1250 for training respectively of the baby cry, glass-break and gunshot models, while we used 100 sequences as validation set and 500 sequences as test set for all of them. In the Evaluation set, the training and test sequences of the Development set are combined into a single training set, while the validation set is the same used in the Development dataset. The system is evaluated against an ``unseen'' set of 1500 samples (500 for each target class) with a sound event presence probability for each class equal to 0.5.

\subsection{Evaluation Metrics}
 
In this work we used the Error Rate (ER) as primary evaluation metric to ensure comparability with the reference systems. In particular, for the evaluations on the TUT-SED 2016 and 2017 datasets we consider a segment-based ER with a one-second segment length, while for the TUT-Rare 2017 the evaluation metric is event-based error rate calculated using onset-only condition with a collar of 500 ms. 
In the segment-based ER the ground truth and system output are compared in a fixed time grid, thus sound events are marked as active or inactive in each segment. For the event-based ER the ground truth and system output are compared at event instance level.

ER score is calculated in a single time frame of one second length from intermediate statistics, i.e., the number of substitutions ($S(t_1)$), insertions ($I(t_1)$), deletions ($D(t_1)$) and active sound events from annotations ($N(t_1)$) for a segment $t_1$. Specifically:
\begin{enumerate}
	
	\item Substitutions $S(t_1)$ are the number of ground truth events for which we have a false positive and one false negative in the same segment; 
	
	\item Insertions $I(t_1)$ are events in system output that are not present in the ground truth, thus the false positives which cannot be counted as substitutions;

	\item Deletions $D(t_1)$ are events in ground truth that are not correctly detected by the system, thus the false negatives which cannot be counted as substitutions.
	
\end{enumerate}
These intermediate statistics are accumulated over the segments of the whole test set to compute the evaluation metric ER. Thus, the total error rate is calculated as:
\begin{equation}
ER = \frac{\sum_{t_1=1}^{T} S(t_1) + \sum_{t_1=1}^{T} I(t_1) + \sum_{t_1=1}^{T} D(t_1)}{\sum_{t_1=1}^{T} N(t_1)},
\end{equation}
where $T$ is the total number of segments $t_1$.

If there are multiple scenes in the dataset, such as in the TUT-SED 2016, evaluation metrics are calculated for each scene separately and then the results are presented as the average across the scenes.
A detailed and visualized explanation of segment-based ER score in multi label setting can be found in \cite{mesaros2016metrics}.

\subsection{Comparative Algorithms}

Since the datasets we used were employed to develop and evaluate the algorithms proposed from the participants of the DCASE challenges, we can compare our results with the most recent approaches in the state-of-the-art. In addition, each challenge task came along with a baseline method that consists in a basic approach for the SED. It represents a reference for the participants of the challenges while they were developing their systems.

\subsubsection{TUT-SED 2016}
The baseline system is based on mel-frequency cepstral coefficients (MFCC) acoustic features and multiple GMM-based classifiers. In detail, for each event class, a binary classifier is trained using the audio segments annotated as belonging to the model representing the event class, and the rest of the audio to the model which represents the negative class. The decision is based on likelihood ratio between the positive and negative models for each individual class, with a sliding window of one second.
To the best of our knowledge, the most performing method for this dataset is an algorithm we proposed \cite{valenti2017neural} in 2017, based on binaural MFCC features and a Multi-Layer Perceptron (MLP) neural network used as classifier. The detection task is performed by an adaptive energy Voice Activity Detector (VAD) which precedes the MLP and determines the starting and ending point of an event-active audio sequence.

\subsubsection{TUT-SED 2017}
In this case the baseline method relies on an MLP architecture using 40 LogMels as audio representation \cite{DCASE2017challenge}. The network is fed with a feature vector comprehending 5-frame as temporal context.
The neural network is composed of two dense layers of 50 hidden units per layer with the 20\% of dropout, while the network output layer contains $K$ sigmoid units (where $K$ is
the number of classes) that can be active at the same time and represent the network prediction of event activity for each context window.
The state-of-the-art algorithm is based on the CRNN architecture \cite{adavanne2017report}. The authors compared both monaural and binaural acoustic features, observing that binaural features in general have similar performance as single channel features on the Development dataset although the best result on the Evaluation dataset is obtained using monaural LogMels as network inputs. According to the authors, this can suggest that the dataset was possibly not large enough to train the CRNN fed with this kind of features.

\subsubsection{TUT-Rare 2017}
The baseline \cite{mesaros2016tut} and the state-of-the-art methods of the DCASE 2017 challenge (Rare-SED) were based on a very similar architectures to that employed for the TUT-SED 2016 and described above. For the baseline method, the only difference relies in the output layer, which in this case is composed of a single sigmoid unit.
The first classified algorithm \cite{limrare} takes 128 LogMels as input and process them frame-wise by means of a CRNN with 1D filters on the first stage.

\subsection{Neural Network configuration}

We performed a hyperparameter search by running a series of experiments over predetermined ranges. We selected the configuration that leads, for each network architecture, to the best results from the cross-validation procedure on the Development dataset of each task and used this architecture to compute the results on the corresponding Evaluation dataset.

The number and shape of convolutional layers, the non-linear activation function, the regularizers in addition to the capsules dimensions and the maximum number of routing iterations have been varied for a total of 100 configurations. Details of searched hyperparameters and their ranges are reported in \tableref{tbl:hyper-params-capsule}.
The neural networks training was accomplished by the AdaDelta stochastic gradient-based optimization algorithm \cite{zeiler2012adadelta} for a maximum of 100 epochs and batch size equal to 20 on the margin loss function. The optimizer hyperparameters were set according to \cite{zeiler2012adadelta} (i.e., initial learning rate $lr=1.0$, $\rho=0.95$, $\epsilon=10^{-6}$). 
The trainable weights were initialized according to the \textit{Glorot-uniform} scheme \cite{glorot2010understanding} and an early stopping strategy was employed during the training in order to avoid overfitting. If the validation ER did not decrease for 20 consecutive epochs, the training was stopped, and the last saved model was selected as the final model. In addition, dropout and $L_2$ weight normalization (with $\lambda=0.01$) have been used as weights regularization techniques \cite{srivastava2014dropout}. 
The algorithm has been implemented in the Python language using Keras \cite{chollet2015keras} and Tensorflow \cite{tensorflow2015-whitepaper} as deep learning libraries, while Librosa \cite{mcfee2015librosa} has been used for feature extraction\footnote{Source code available at the following address: \url{https://gitlab.com/a3labShares/capsule-for-sed}}.

 \begin{table}[!t]
	\centering
	\caption{Hyperparameters optimized in the random-search phase and the resulting best performing models.}		
	\label{tbl:hyper-params-capsule}
	\begin{tabular} { lc c}
		\toprule
		Parameter & Range & Distribution\\  
		\midrule
		Batch Normalization  & [yes - no]	& random choice  \\
		\midrule
		
		CNN layers Nr.  & [1 - 4]& uniform \\

		CNN kernels Nr. & [4 - 64]& log-uniform \\
		  
		CNN kernels dim. & [3$\times$3 - 8$\times$8]& uniform \\

		Pooling dim. & [1$\times$1 - 2$\times$5] & uniform 	\\

		CNN activation & [tanh - ReLU] & random choice \\
		
		CNN dropout  & [0 - 0.5]	& uniform  \\
		
		CNN L2  & [yes - no]	& random choice  \\		
		
		\midrule
		Primary Capsules Nr. $M$ & [2 - 8]	& uniform  \\

		Primary Capsules kernels dim. & [3$\times$3 - 5$\times$5]& uniform  \\

		Primary Capsules dimension $J$ & [2 - 16]	& uniform \\

		Detection Capsules dimension $G$ & [2 - 16]	& uniform \\
		
		Capsules dropout  & [0 - 0.5]	& uniform  \\
		
		Routing iterations  & [1 - 5]	& uniform \\
	
		\bottomrule
	\end{tabular}
\end{table}
 
For the CNN models, we performed a similar random hyperparameters search procedure for each dataset, considering only the first two blocks of the \tableref{tbl:hyper-params-capsule} and by replacing the capsule layers with feedforward layers with \textit{sigmoid} activation function. 

On TUT-SED 2016 and 2017 datasets, the event activity probabilities are simply thresholded at a fixed value equal to 0.5, in order to obtain the binary activity matrix used to compute the reference metric. On the TUT-Rare 2017 the network output signal is processed as proposed in \cite{vesperini2018hierarchic}, thus it is convolved with an exponential decay window then it is processed with a sliding median filter with a local window-size and finally a threshold is applied.

\section{Results}
\label{sec:results}

In this section, we present the results for all the datasets and experiments described in \secref{sec:experiment}. The evaluation of Capsule and CNNs based methods have been conducted on the Development sets of each examined dataset using random combinations of hyperparameters given in \tableref{tbl:hyper-params-capsule}. 

\subsection{TUT-SED 2016}
Results on TUT-SET 2016 dataset are shown in \tableref{tbl:results-dcase2016}, while \tableref{tbl:params-dcase2016} reports the configurations which yielded the best performance on the Evaluation dataset. 
All the found models have ReLU as non-linear activation function and use dropout technique as weight regularization, while the batch-normalization applied after each convolutional layer seems to be effective only for the CapsNet. \tableref{tbl:results-dcase2016} reports the results considering each combination of architecture and features we evaluated. The best performing setups are highlighted with bold face. The use of STFT as acoustic representation is beneficial for both the architectures with respect to the LogMels. In particular, the CapsNet obtains the lowest ER on the cross-validation performed on Development dataset when is fed by the binaural version of such features. On the two scenarios of the Evaluation dataset, a model based on CapsNet and binaural STFT obtains an averaged ER equal to 0.69, which is largely below both the challenge baseline \cite{mesaros2016tut} (-0.19) and the best score reported in literature \cite{valenti2017neural} (-0.10). 
The comparative method based on CNNs seems not to fit at all when LogMels are used as input, while the performance is aligned with the challenge baseline based on GMM classifiers when the models are fed by monaural STFT. This discrepancy can be motivated by the enhanced ability of CapsNet to exploit small training datasets, in particular due to the effect of the routing mechanism on the weight training. In fact, the TUT-SED 2016 is composed of a small amount of audio and the sounds events occur sparsely (i.e., only 49 minutes of the total audio contain at least one event active), thus, the overall results of the comparative methods (CNNs, Baseline, and State-of-the-art) on this dataset are quite low compared to the other datasets. 

Another CapsNet property that is worth to highlight is the lower number of free parameters that compose the models compared to evaluated CNNs.
As shown in \tableref{tbl:params-dcase2016}, the considered architectures have 267\,K and 252\,K free parameters respectively for the ``Home'' and the ``Residential area'' scenario. It is a relatively low number of parameters to be trained (e.g., a popular deep architecture for image classification such as AlexNet \cite{krizhevsky2012imagenet} is composed of 60\,M parameters), and the best performing CapsNets of each considered scenario have even less parameters with respect to the CNNs (-22\% and -64\%  respectively for the ``Home'' and the ``Residential area'' scenario). Thus, the high performance of CapsNet can be explained with the architectural advantage rather than the model complexity. In addition, there can be a significant performance shift for the same type of networks with the same number of parameters,
which means that a suitable hyperparameters search action (e.g., number of filters on the convolutional layers, dimension of the capsule units) is crucial in finding the best performing network structure.

\begin{figure}[t]
	\centering
	\includegraphics[width=0.9\columnwidth]{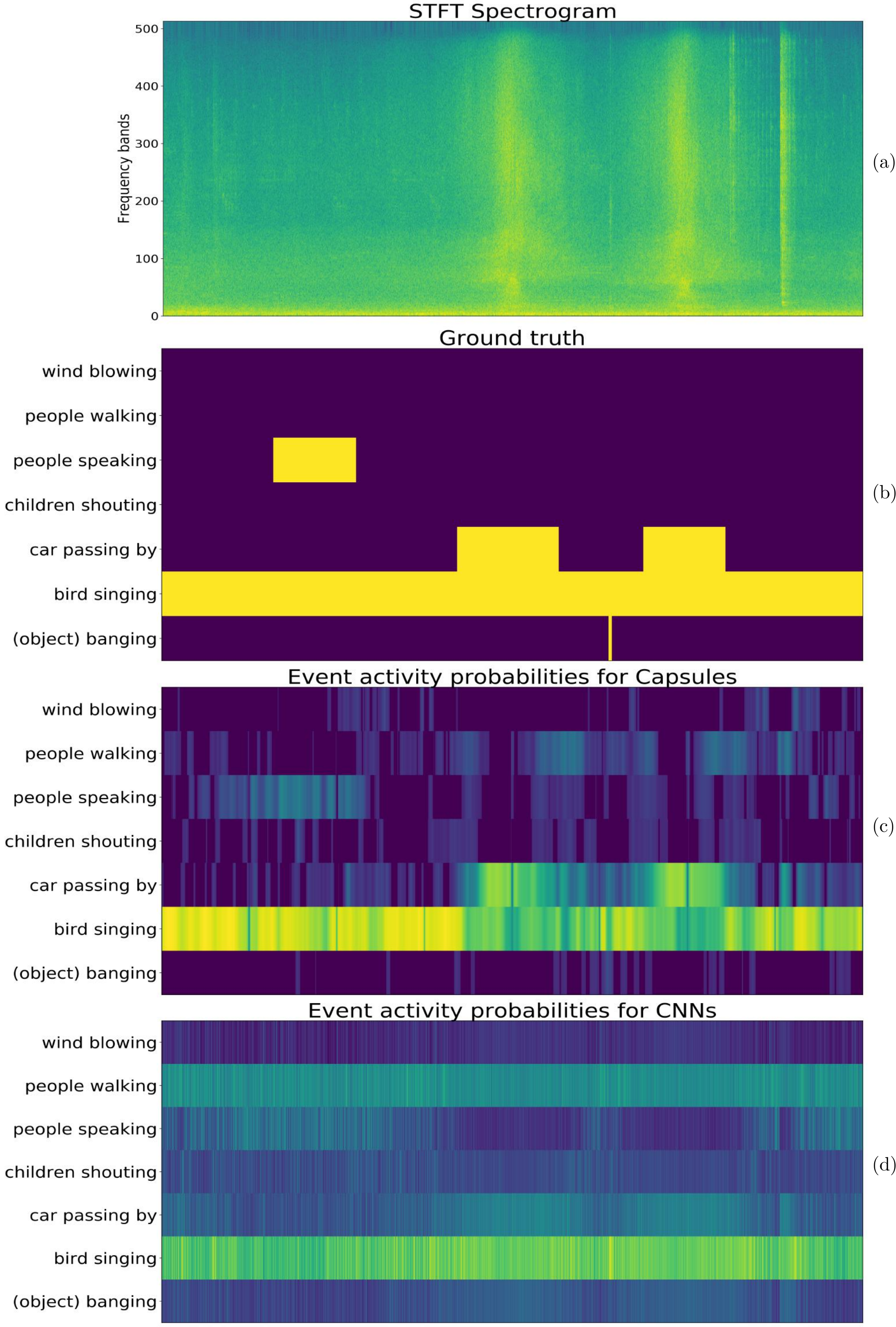}
	\caption{STFT Spectrogram of the input sequence (a), ground truth (b) and event activity probabilities for CapsNet (c) and CNN (d) from a sequence of test examples from TUT-SED 2016 dataset.}
	\label{fig:activations}
\end{figure}

\subsubsection{Closer Look at Network Outputs}
A \textcolor{black}{comparative example} on the neural network outputs, which are regarded as event activity probabilities is presented in \figref{fig:activations}. The monaural STFT from a 40 seconds sequence of the ``Residential area'' dataset is shown along with event annotations and the network outputs of the CapsNet and the CNN best performing models. For this example, we chose the monaural STFT as input feature because generally it yields the best results over all the considered datasets. \figref{fig:activations} shows a \textit{bird singing} event lasting for the whole sequence and correctly detected by both the architectures. When the \textit{car passing by} event overlaps the \textit{bird singing}, the CapsNet detects more clearly its presence. The \textit{people speaking} event is slightly detected by both the models, while the \textit{object banging} activates the relative Capsule exactly only in correspondence of the event annotation. It must be noted that the dataset is composed of unverified manually labelled real-life recordings, that may present a degree of subjectivity, thus, affecting the training. Nevertheless, the CapsNet exhibits remarkable detection capability especially in the condition of overlapping events, while the CNN outputs are definitely more ``blurred'' and the event \textit{people walking} is wrongly detected in this sequence.

\begin{table*}[!t]
	\centering
	\caption{Hyperparameters of the best performing models on the TUT-Polyphonic SED 2016 \& 2017 Evaluation datasets.}		
	\label{tbl:params-dcase2016}
	\begin{tabular}{@{}lllllll@{}}
		\toprule
		& \multicolumn{4}{c}{\textbf{TUT-SED 2016}}                                                                     & \multicolumn{2}{c}{\textbf{TUT-SED 2017}}             \\ \midrule
		& \multicolumn{2}{c}{Home}                              & \multicolumn{2}{c}{Residential}                       & \multicolumn{2}{c}{Street}                            \\ \midrule
		& \multicolumn{1}{c}{CapsNet} & \multicolumn{1}{c}{CNN} & \multicolumn{1}{c}{CapsNet} & \multicolumn{1}{c}{CNN} & \multicolumn{1}{c}{CapsNet} & \multicolumn{1}{c}{CNN} \\ \cmidrule(l){2-7} 
		CNN kernels Nr.               & $[32, 32, 8]$               & $[64,64,16,64]$         & $[4,16,32,4]$               & $[64]$                  & $[4,16,32,4]$               & $[64, 64,16, 64]$       \\
		CNN kernels dim.              & $6\times6$                  & $5\times5$              & $4\times4$                  & $5\times5$              & $4\times4$                  & $5\times5$              \\
		Pooling dim. ($F$ axis)       & $[4,3,2]$   				& $[2,2,2,2]$             & $[2,2,2,2]$                 & $[2]$             	  & $[2,2,2,2]$                 & $[2,2,2,2]$  			  \\
		MLP layers dim.               & -                           & $[85, 65]$              & -                           & $[42, 54, 66, 139]$     & -                           & $[85, 65]$              \\ \midrule
		Primary Capsules Nr. $M$ & 8                           & -				          & 7                           & -                       & 7                           & -                       \\
		Primary Capsules kernels dim. & $4\times4$                  & -                       & $3\times3$                  & -                       & $3\times3$                  & -                       \\
		Primary Capsules dimension   $J$ & 9                           & -                       & 16                          & -                       & 16                          & -                       \\
		Detection Capsules dimension         $G$   & 11                          & -                       & 8                           & -                       & 8                           & -                       \\
		Routing iterations            & 3                           & -                       & 4                           & -                       & 4                           & -                       \\
		\midrule
		\# Params                     & 267\,K                        & 343\,K                    & 252\,K                        & 709\,K                    & 223K                        & 342\,K                    \\ \bottomrule
	\end{tabular}
\end{table*}

\begin{table*}[!t]
	\centering
	\caption{Results of best performing models in terms of ER on the TUT-SED 2016 \& 2017 dataset.}		
	\label{tbl:results-dcase2016}
	\begin{tabular}{@{}lllllllll@{}}
		\toprule
		\multicolumn{9}{c}{\textbf{TUT-SED 2016 - Home}}                                                                                         \\ \midrule
		\multicolumn{5}{c|}{Development}                                                           & \multicolumn{4}{c}{Evaluation}                        \\ \midrule
		Features      & LogMels & Binaural  LogMels & STFT & \multicolumn{1}{l|}{Binaural STFT} & LogMels & Binaural LogMels & STFT & Binaural STFT \\ \midrule
		CNN           & 11.15     & 11.58              & 1.06 & \multicolumn{1}{l|}{1.07}        & 6.80      & 8.43               & 0.95 & 0.92       \\
		CapsNet      & 0.58      & 0.59               & 0.44 & \multicolumn{1}{l|}{\textbf{0.39}}        & 0.74      & 0.75               &\textbf{0.61} & 0.69        \\ \midrule
		\multicolumn{9}{c}{\textbf{TUT-SED 2016 - Residential Area}}                                                                             \\ \midrule
		Features      & LogMels & Binaural  LogMels & STFT & \multicolumn{1}{l|}{Binaural STFT} & LogMels & Binaural LogMels & STFT & Binaural STFT \\ \midrule
		CNN           & 3.24      & 3.11               & 0.64 & \multicolumn{1}{l|}{1.10}        & 2.36      & 2.76               & 1.00 & 1.35        \\
		CapsNet       & 0.36      & 0.34               & 0.32 & \multicolumn{1}{l|}{\textbf{0.32}}        & 0.72      & 0.75               & 0.78 &\textbf{0.68}        \\ \midrule
		\midrule
		\multicolumn{9}{c}{\textbf{TUT-SED 2016 - Averaged}}                                                                                         \\ \midrule
		CNN     & 7.20      & 7.35               & 0.85 & \multicolumn{1}{l|}{1.09}        & 4.58      & 5.60               & 0.98 & 1.14       \\
		CapsNet & 0.47      & 0.47               & 0.38 & \multicolumn{1}{l|}{\textbf{0.36}}        & 0.73      & 0.75               & 0.70 & \textbf{0.69}       \\
		\midrule
		Baseline \cite{mesaros2016tut}      & \multicolumn{4}{c|}{0.91}                                                & \multicolumn{4}{c}{0.88}                            \\
		State-of-the-art \cite{valenti2017neural}          & \multicolumn{4}{c|}{0.78}                                                & \multicolumn{4}{c}{0.79}                            \\ \bottomrule
		\toprule
		\multicolumn{9}{c}{\textbf{TUT-SED 2017}}                                                                                                                                  \\ \midrule
		& \multicolumn{4}{c|}{Development}                                                          & \multicolumn{4}{c}{Evaluation}                                      \\ \midrule
		Features & LogMels & Binaural  LogMels & STFT & \multicolumn{1}{l|}{Binaural STFT} & LogMels & Binaural LogMels & STFT & Binaural STFT \\ \midrule
		CNN      & 1.56             & 2.12                       & 0.57 & \multicolumn{1}{l|}{0.60}          & 1.38             & 1.79                      & 0.67 & 0.65        \\
		CapsNet  & 0.45             & 0.42                       & \textbf{0.36} & \multicolumn{1}{l|}{0.36} &\textbf{0.58}            & 0.64                      & 0.61 & 0.64 \\ 
		\midrule
		Baseline \cite{DCASE2017challenge} & \multicolumn{4}{c|}{0.69}							& \multicolumn{4}{c}{0.93} \\
		State-of-the-art \cite{adavanne2017report}     & \multicolumn{4}{c|}{0.52}             				& \multicolumn{4}{c}{0.79}  \\ \bottomrule
	\end{tabular}
\end{table*}

\begin{figure}[t]
	\centering
	\includegraphics[width=0.9\columnwidth]{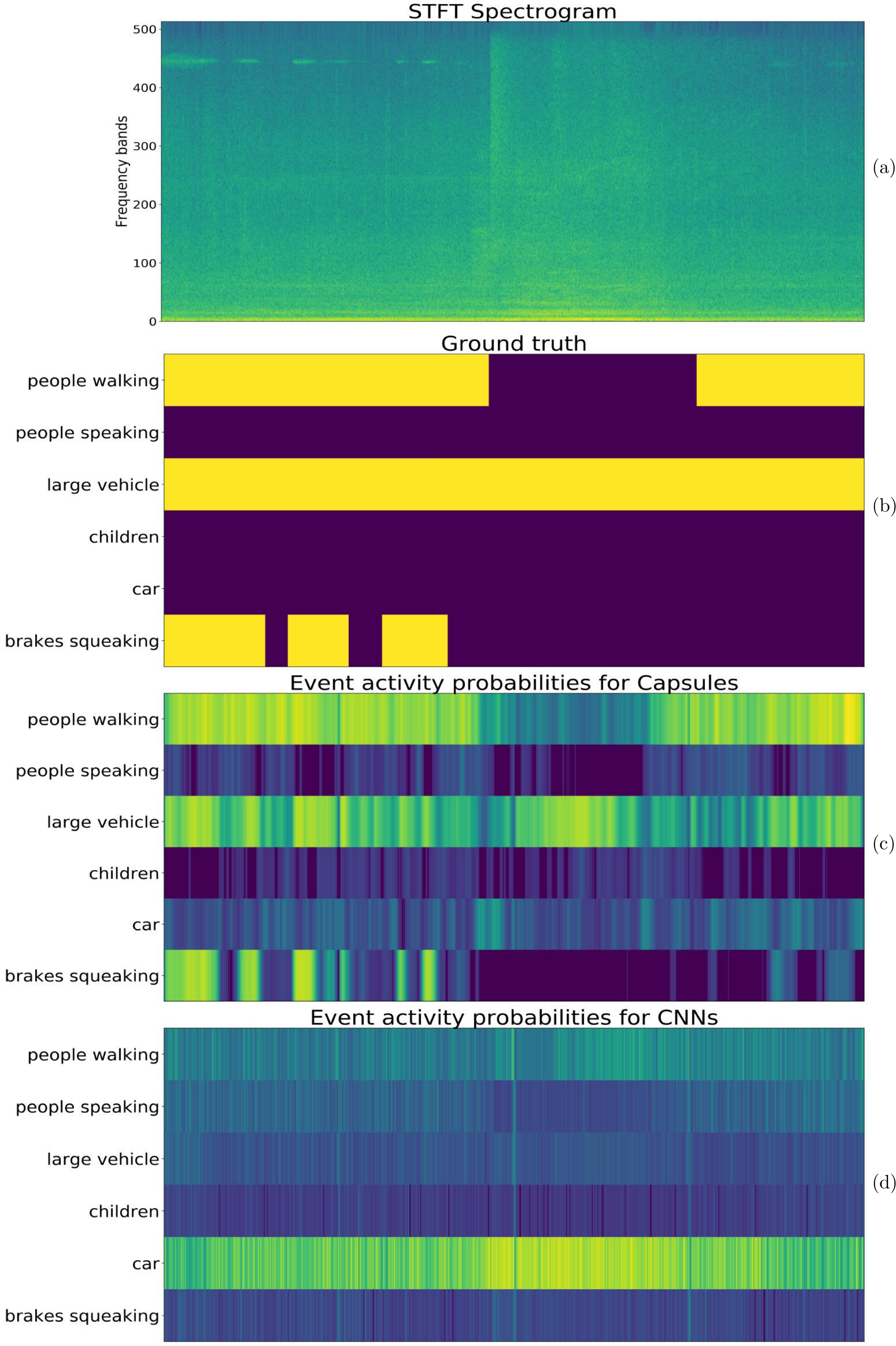}
	\caption{STFT Spectrogram of the input sequence (a), ground truth (b) and event activity probabilities for CapsNet (c) and CNN (d) from a sequence of test examples from TUT-SED 2017 dataset.}
	\label{fig:activations_17}
\end{figure}

\subsection{TUT-SED 2017}

The bottom of \tableref{tbl:results-dcase2016} reports the results obtained with the TUT-SED 2017. As in the TUT-SED 2016, the best performing models on the Development dataset are those fed by the Binaural STFT of the input signal. In this case, we can also observe interesting performance obtained by the CNNs, which on the Evaluation dataset obtain a lower ER (i.e., equal to 0.65) with respect to the state-of-the-art algorithm \cite{adavanne2017report}, based on CRNNs. CapsNet confirms its effectiveness and it obtains lowest ER equal to 0.58 with LogMel features, although with a slight margin with respect to the other inputs (i.e., -0.03 compared to the STFT features, -0.06 compared to both the binaural version of LogMels and STFT spectrograms).

It is worth highlighting that in the Development cross-validation, the CapsNet models yielded significantly better performance with respect to the other reported approaches, while the CNNs have decidedly worse performance. {\color{black}On the Evaluation dataset, however, the ER scores of the CapsNets suffer more relative deterioration with respect to the CNNs ones. This is related to the fact that the CapsNet are subject to larger random fluctuations of the ER from epoch to epoch. In absence of ground truth labels and, thus, of the early stopping strategy, the model taken after a fixed number of training epochs is suboptimal, and, with CapsNet, more prone to large errors than with CNN.}

Notwithstanding this weakness, the absolute performance obtained both with monaural and binaural spectral features is consistent and improves the state-of-the-art result, with a reduction of the ER of up to 0.21 in the best case. This is particularly evident in \figref{fig:activations_17}, that shows the output of the two best performing systems for a sequence of approximately 20 seconds length which contains highly overlapping sounds. The event classes ``people walking'' and ``large vehicle'' are overlapped for almost all the sequence duration and they are well detected by the CapsNet, although they are of different nature: the ``large vehicle'' has a typical timber and is almost stationary, while the class ``people walking'' comprehend impulsive and desultory sounds. The CNN does not seem to be able to distinguish between the ``large vehicle'' and the ``car'' classes, detecting confidently only the latter, while the activation corresponding ``people walking'' class is modest. The presence of the ``brakes squeaking'' class, which has a specific spectral profile mostly located in the highest frequency bands (as shown in the spectrogram), is detected only by the CapsNet. We can assume this as a concrete experimental validation of the routing effectiveness.
 
The number of free parameters amounts to 223\,K for the best configuration shown in \tableref{tbl:params-dcase2016} and it is similar to those found for the TUT-SED 2016, which consists also in this case in a reduction equal to 35\% with respect to the best CNN layout.


\begin{table*}[ht]
	\centering
	\caption{Hyperparameters of the best performing models on the TUT-Rare 2017 Monophonic SED Evaluation datasets.}		
	\label{tbl:params-dcase2017-rare}
	\begin{tabular}{@{}lllllll@{}}
		\toprule
		\multicolumn{7}{c}{\textbf{TUT-Rare SED 2017 Monophonic SED}}                                                                 \\ \midrule
		& \multicolumn{2}{c}{Baby cry}   & \multicolumn{2}{c}{Glass break} & \multicolumn{2}{c}{Gunshot}  \\ \midrule
		& CapsNet      & CNN            & CapsNet       & CNN            & CapsNet    & CNN             \\ \cmidrule(l){2-7} 
		CNN kernels Nr.               & $[16,64,32]$ & $[16,32,8,16]$ & $[16,64,32]$  & $[16,32,8,16]$ & $[16,16]$  & $[16,64,32,32]$ \\
		CNN kernels dim.              & $6\times6$   & $8\times8$     & $6\times6$    & $8\times8$     & $8\times8$ & $7\times7$      \\
		Pooling dim. ($F$ axis)       & $[4,3,2]$    & $[3,3,2,2]$    & $[4,3,2]$     & $[3,3,2,2]$    & $[5,2]$    & $[5,4,2,1]$     \\
		MLP layers dim.               & -            & $[212,67]$     & -             & $[212,67]$     & -          & $112,51$        \\ \midrule
		Primary Capsules Nr. $M$& 7            & -               & 7             & -               & 8          & -                \\
		Primary Capsules kernels dim.  & $3\times3$   & -               & $3\times3$    & -               & $3\times3$ & -                \\
		Primary Capsules dimension  $J$  & 8            & -               & 8             & -               & 8          & -                \\
		Detection Capsules dimension      $G$      & 14           & -               & 14            & -               & 6          & -                \\
		Routing iterations            & 5            & -               & 5             & -              & 1          & -                \\ \midrule
		\# Params                     & 131\,K         & 84\,K            & 131\,K          & 84\,K            & 30\,K        & 211\,K            \\ \bottomrule
	\end{tabular}
\end{table*}

\subsection{TUT-Rare SED 2017}
The advantage provided by the routing procedure to the CapsNet is particularly effective in the case of polyphonic SED. The results on the monophonic SED task have been obtained by using the TUT-Rare SED 2017 dataset and they are shown in \tableref{tbl:results-dcase2017Rare}. In this case, the evaluation metric is the event-based ER calculated using onset-only condition. We performed a separate random-search for each of the three sound event classes both for CapsNets and CNNs and we report the averaged score over the three classes. The setups that obtained the best performance on the Evaluation dataset are shown in \tableref{tbl:params-dcase2017-rare}. This is the largest dataset we evaluated, and its characteristic is the high unbalance between the amount of background sounds versus the target sound events.

From the analysis of the results of the individual classes on the Evaluation set (not included here for the sake of conciseness), we notice that both architectures achieve the best performance on the \textit{glass break} class (0.25 and 0.24 respectively for CNNs and CapsNet with LogMels features), due to its clear spectral fingerprint compared to the background sound. The worst performing class is the \textit{gunshot} (ER equal to 0.58 for the CapsNet), although the noise produced by different instances of this class involves similar spectral components. The low performance is probably due to the fast decay of this sound, which means that in this case the routing procedure is not sufficient to avoid confusing the \textit{gunshot} with other background noises, especially in the case of dataset unbalancing and low event-to-background ratio. A solution to this issue can be found in the combination of CapsNet with RNN units, as proposed in \cite{cakir2017convolutional} for the CNNs which yields an efficient modelling of the \textit{gunshot} by CRNN and improves the detection abilities even in polyphonic conditions. The \textit{baby cry} class consists of short, harmonic sounds, and it is detected with almost the same accuracy by the two architectures.

Finally, the CNN shows better generalization performance with respect to the CapsNet, although the ER score is far from state-of-the-art that use the aforementioned CRNNs \cite{limrare} or a hierarchical framework \cite{vesperini2018hierarchic}. In addition, in this case the CNN models have a reduced number of trainable parameters (36\%) compared to the CapsNets, except for the ``gunshot'' case but, as mentioned, it is also the configuration that gets the worst results.

\begin{table}[t]
	\centering 
	\caption{Results of best performing models in terms of ER on the TUT-RareSED 2017 dataset.}		
	\label{tbl:results-dcase2017Rare}
	\begin{tabular}{@{}lllll@{}}
		\toprule
		\multicolumn{5}{c}{\textbf{TUT-RareSED 2017 - Monophonic SED}}                           \\ \midrule
		& \multicolumn{2}{c|}{Development}             & \multicolumn{2}{c}{Evaluation} \\ \midrule
		Features & LogMels & \multicolumn{1}{l|}{STFT} & LogMels     & STFT    \\
		CNN      & 0.29             & \multicolumn{1}{l|}{0.21} & \textbf{0.41}        & 0.46    \\
		CapsNet  & \textbf{0.17}    & \multicolumn{1}{l|}{0.20} & 0.45                 & 0.54    \\
		\hline
		Baseline \cite{mesaros2016tut} & 0.53             & \multicolumn{1}{l|}{}     & 0.64                &         \\	
		Hierarchic CNNs  \cite{vesperini2018hierarchic}   & 0.13             & \multicolumn{1}{l|}{}     & 0.22                 &         \\
		State-of-the-art  \cite{limrare}    &\textbf{0.07}             & \multicolumn{1}{l|}{}     &\textbf{0.13}                 &         \\ \bottomrule
	\end{tabular}
	\end{table}

\begin{table}[t]
	\centering 
	\caption{Results of test performed with our proposed variant of routing procedure.}		
	\label{tbl:results-new-routing}
	\begin{tabular}{@{}lllll@{}}
		\toprule
		\multicolumn{5}{c}{\textbf{TUT-SED 2016 - Home}}                                   \\ \midrule
		& \multicolumn{2}{c|}{Development}   & \multicolumn{2}{c}{Evaluation} \\ \midrule
		CapsNet      & 0.44 & \multicolumn{1}{l|}{}       & 0.61              &            \\
		CapsNet - NR & 0.41 & \multicolumn{1}{l|}{-6.8\%} & \textbf{0.58}     & -4.9 \%    \\ \midrule
		\multicolumn{5}{c}{\textbf{TUT-SED 2016 - Residential}}                            \\ \midrule
		CapsNet      & 0.32 & \multicolumn{1}{l|}{}       & 0.78              &            \\
		CapsNet - NR & 0.31 & \multicolumn{1}{l|}{-3.1\%} & \textbf{0.72}     & -7.7 \%    \\ \midrule
		\multicolumn{5}{c}{\textbf{TUT-SED 2016 - Average}}                                \\ \midrule
		CapsNet      & 0.38 & \multicolumn{1}{l|}{}       & 0.70              &            \\
		CapsNet - NR & 0.36 & \multicolumn{1}{l|}{-5.3\%} & \textbf{0.65}     & -7.1 \%    \\ \midrule
		\multicolumn{5}{c}{\textbf{TUT-SED 2017 - Street}}                                 \\ \midrule
		CapsNet      & 0.36 & \multicolumn{1}{l|}{}       & 0.61              &            \\
		CapsNet - NR & 0.36 & \multicolumn{1}{l|}{0.0\%} & \textbf{0.58}     & -4.9 \%    \\ \bottomrule
	\end{tabular}
\end{table}

\subsection{Alternative Dynamic Routing for SED}
We observed that the original routing procedure implies the initialization of the coefficients $\beta_{ij}$ to zero each time the procedure is restarted, i.e., after each input sample has been processed. This is reasonable in the case of image classification, for which the CapsNet has been originally proposed. In the case of audio task, we clearly expect a higher correlation between samples belonging to adjacent temporal frames $\mathbf{X}$. We thus investigated the chance to initialize the coefficients $\beta_{ij}$ to zero only at the very first iteration, while for subsequent $\mathbf{X}$ to assign them the last values they had at the end of the previous iterative procedure. We experimented this variant considering the best performing models of the analyzed scenarios for polyphonic SED, taking into account only the systems fed with the monaural STFT. As shown in \tableref{tbl:results-new-routing}, the modification we propose in the routing procedure is effective in particular on the Evaluation datasets, conferring improved generalization properties to the models we tested even without accomplishing a specific hyperparameters optimization. 

\section{Conclusion}
\label{sec:conclusions}
In this work, we proposed to apply a novel neural network architecture, the CapsNet, to the polyphonic SED task. The architecture is based on both convolutional and capsule layers. The convolutional layers extract high-level time-frequency feature maps from input matrices which provide an acoustic spectral representation with long temporal context. The obtained feature maps are then used as input to the Primary Capsule layer which is connected to the Detection Capsule layer that extracts the event activity probabilities. These last two layers are involved in the iterative \textit{routing-by-agreement} procedure, which computes the outputs based on a measure of likelihood between a capsule and its parent capsules. This architecture combines, thus, the ability of convolutional layers to learn local translation invariant filters with the ability of capsules to learn part-whole relations by using the routing procedure. 
 
Part of the novelty of this work resides in the adaptation of the CapsNet architecture for the audio event detection task, with a special care on the input data, the layers interconnection and the regularization techniques. The routing procedure is also modified to \textcolor{black}{account for an assumed temporal correlation within the data}, with a further average performance improvement of 6\% among the polyphonic SED tasks.

An extensive evaluation of the algorithm is proposed with comparison to recent state-of-the-art methods on three different datasets. The experimental results demonstrate that the use of dynamic routing procedure is effective and it provides significant performance improvement in the case of overlapping sound events compared to traditional CNNs, and other established methods in polyphonic SED. Interestingly, the CNN based method obtained the best performance in the monophonic SED case study, thus emphasizing the suitability of the CapsNet architecture in dealing with overlapping sounds.
We showed that this model is particularly effective with small sized datasets, such as TUT-SED 2016 which contains a total 78 minutes of audio for the development of the models of which one third is background noise. Furthermore, the network trainable parameters are reduced with respect to other deep learning architectures, confirming the architectural advantage given by the introduced features also in the task of polyphonic SED.

The results we observed in this work are consistent with many other classification tasks in various domains \cite{deng2018hyper, shen2018dynamic, jalal2018american} and they prove that the CapsNet is an effective approach which enhances the well-established representation capabilities of the CNNs also in the audio field. \textcolor{black}{   
    However, several aspects still remain unexplored and require further studies: the robustness of CapsNets to overlapping signals (i.e., images or sounds) has been demonstrated in this work as well as in \cite{sabour2017dynamic}. In \cite{sabour2017dynamic}, the authors demonstrated also the capability of CapsNets to be invariant to affine transformations of images, such as rotations. In the audio case study, this characteristic could be exploited for obtaining invariance respect to the source position by using a space-time representation of multi-channel audio signals.    
} Moreover, regularization methods can be investigated to overcome the lack of generalization which seems to affect the CapsNets. Furthermore, regarding the SED task the addition of recurrent units may be explored to enhance the detection of particular (i.e., impulsive) sound events in real-life audio and the recently-proposed variant of routing, based on the Expectation Maximization algorithm (EM) \cite{hinton2018matrix}, can be investigated in this context.


%


\section*{Acknowledgement}
This research has been partly supported by the Italian University and Research Consortium CINECA. We acknowledge them for the availability of high-performance computing resources and support.

\ifCLASSOPTIONcaptionsoff
  \newpage
\fi



%

\bibliographystyle{IEEEtran}
\bibliography{refs}

%

\begin{IEEEbiography}[{\includegraphics[width=1in,height=1.25in,clip,keepaspectratio]{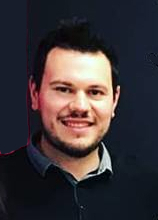}}]
	{Fabio Vesperini} was born in San Benedetto del Tronto, Italy, on May 1989. He received his M.Sc. degree (cum laude) in electronic engineering in 2015 from Universit\`{a} Politecnica delle Marche (UnivPM). In 2014 he was at the Technische Universtit\"at M\"unchen as visiting student for 7 months, where he carried out his master thesis project on acoustic novelty detection. He is currently a PhD student at the Department of Information Engineering, at UnivPM. His research interests are in the fields of digital signal processing and machine learning for intelligent audio analysis.
\end{IEEEbiography}

\begin{IEEEbiography}[{\includegraphics[width=1in,height=1.25in,clip,keepaspectratio]{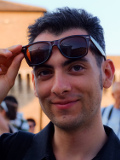}}]
	{Leonardo Gabrielli} got his M.Sc. and PhD degrees in Electronics Engineering from Università Politecnica delle Marche, Italy, respectively in 2011 and 2015. His main research topics are related to audio signal processing and machine learning with application to sound synthesis, Computational Sound Design, Networked Music Performance, Music Information Retrieval and audio classification. He has been co-founder of DowSee srl, and holds several industrial patents. He is coauthor of more than 30 scientific papers.
\end{IEEEbiography}

\begin{IEEEbiography}[{\includegraphics[width=1in,height=1.25in,clip,keepaspectratio]{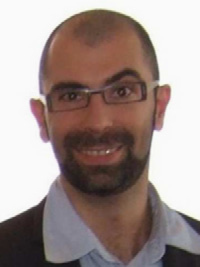}}]
	{Emanuele Principi} was born in Senigallia (Ancona), Italy, on January 1978. He received the M.S. degree in electronic engineering (with honors) from Universit\`a Politecnica delle Marche (Italy) in 2004. He received his Ph.D. degree in 2009 in the same university under the supervision of Prof. Francesco Piazza. In November 2006 he joined the 3MediaLabs research group coordinated by Prof. Francesco Piazza at Universit\`a Politecnica delle Marche where he collaborated to several regional and european projects on audio signal processing. Dr. Principi is author and coauthor of several international scientific peer-reviewed articles in the area of speech enhancement for robust speech and speaker recognition and intelligent audio analysis. He is member of the IEEE CIS Task Force on Computational Audio Processing, and is reviewer for several international journals. His current research interests are in the area of machine learning and digital signal processing for the smart grid (energy task scheduling, non-intrusive load monitoring, computational Intelligence for vehicle to grid) and intelligent audio analysis (multi-room voice activity detection and speaker localization, acoustic event detection, fall detection).
\end{IEEEbiography}

\begin{IEEEbiography}[{\includegraphics[width=1in,height=1.25in,clip,keepaspectratio]{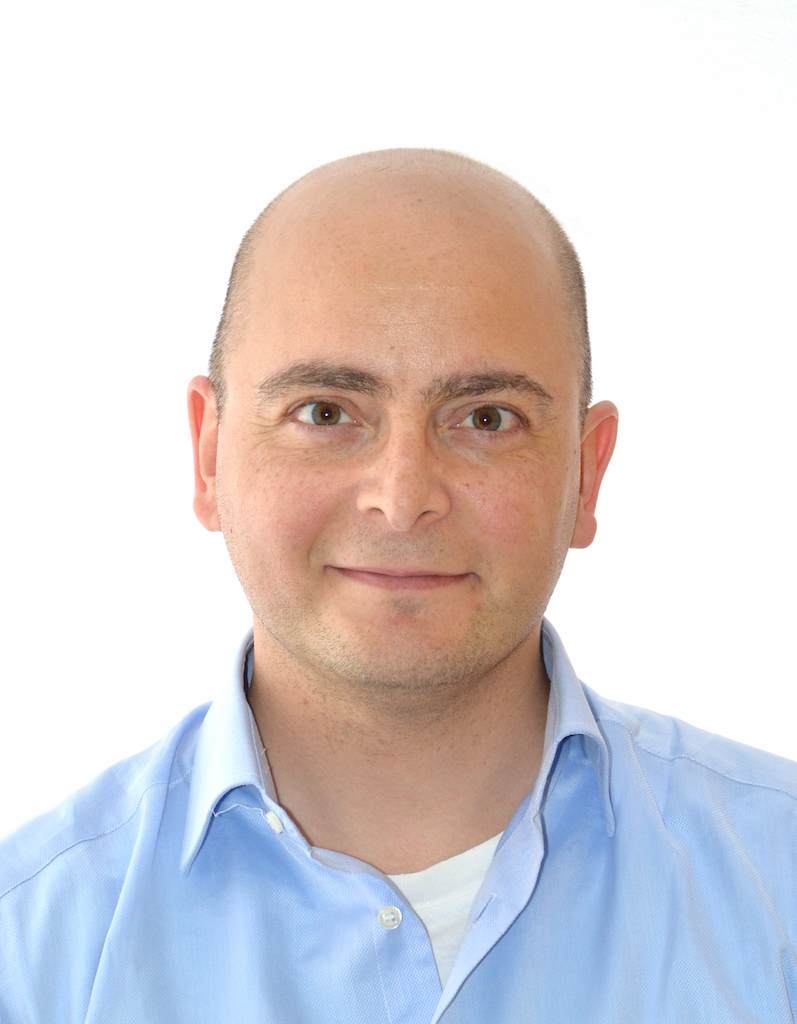}}]{Stefano Squartini} (IEEE Senior Member, IEEE CIS Member) was born in Ancona, Italy, on March 1976. He got the Italian Laurea with honors in electronic engineering from University of Ancona (now Polytechnic University of Marche, UnivPM), Italy, in 2002. He obtained his PhD at the same university (November 2005). He worked also as post-doctoral researcher at UnivPM from June 2006 to November 2007, when he joined the DII (Department of Information Engineering) as Assistant Professor in Circuit Theory. He is now Associate Professor at UnivPM since November 2014. His current research interests are in the area of computational intelligence and digital signal processing, with special focus on speech/audio/music processing and energy management. He is author and coauthor of more than 190 international scientific peer-reviewed articles. He is Associate Editor of the IEEE Transactions on Neural Networks and Learning Systems, IEEE Transactions on Cybernetics and IEEE Transactions on Emerging Topics in Computational Intelligence, and also member of Cognitive Computation, Big Data Analytics and Artificial Intelligence Reviews Editorial Boards. He joined the Organizing and the Technical Program Committees of more than 70 International Conferences and Workshops in the recent past. He is the Organizing Chair of the IEEE CIS Task Force on Computational Audio Processing.
\end{IEEEbiography}



\end{document}